\documentclass{article}

\usepackage{microtype}
\usepackage{graphicx}
\usepackage{subfigure}
\usepackage{booktabs} 
\usepackage{subfiles}
\usepackage{graphicx}
\usepackage{caption}

\usepackage[hyphens]{url}
\usepackage{hyperref}
\hypersetup{breaklinks=true}
\usepackage[accepted]{icml2019}

\icmltitlerunning{Using AI for Economic Upliftment of Handicraft Industry}
\graphicspath{ {./images/} }
\begin{document}

\twocolumn[
\icmltitle{Using AI for Economic Upliftment of Handicraft Industry}



\icmlsetsymbol{equal}{*}

\begin{icmlauthorlist}
\icmlauthor{Nitya Raviprakash}{mi}
\icmlauthor{Sonam Damani}{mi}
\icmlauthor{Ankush Chatterjee}{mi}
\icmlauthor{Meghana Joshi}{mi}
\icmlauthor{Puneet Agrawal}{mi}
\end{icmlauthorlist}

\icmlaffiliation{mi}{Microsoft, Hyderabad, Telangana, India}

\icmlcorrespondingauthor{Sonam Damani}{sodamani@microsoft.com}

\icmlkeywords{Machine Learning, ICML}

\vskip 0.3in
]



\printAffiliationsAndNotice{}  

\begin{abstract}

The handicraft industry is a strong pillar of Indian economy which provides large-scale employment opportunities to artisans in rural and underprivileged communities. However, in this era of globalization, diverse modern designs have rendered traditional designs old and monotonous, causing an alarming decline of handicraft sales. For this age-old industry to survive the global competition, it is imperative to integrate contemporary designs with Indian handicrafts. In this paper, we use novel AI techniques to generate contemporary designs for two popular Indian handicrafts - Ikat and Block Print. These techniques were successfully employed by communities across India to manufacture and sell products with greater appeal and revenue. The designs are evaluated to be significantly more likeable and marketable than the current designs used by artisans.


\end{abstract}

\section{Introduction}

\begin{figure}[t!]
  \centering
  \includegraphics[width=0.45\textwidth]{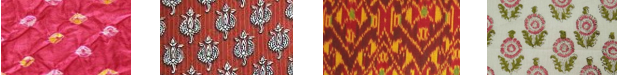}
  \caption{\small{Traditional Indian handcrafted designs}}
  \label{fig:handicrafts}
\end{figure}

For years, the handicrafts industry has largely contributed to employment in rural India. As a highly labor intensive industry, more than 7 million artisans have been employed, a majority of whom are women or belong to underprivileged sections of society. However, the handicrafts industry, which was traditionally a major source of revenue generation in rural India, has suffered severe economic decline in recent years. According to the United Nations, the number of Indian artisans reduced by 30\% over the past 30 years \citep{dasra2013handicrafts}. As a result of globalization, local artisans lack the ability to cater their handicrafts to new markets and are forced to find low unskilled employment in urban industries \citep{harita2014design}. A major factor contributing to the market loss is the continued usage of age-old motifs, shapes, and color schemes by local artisans, in contrast to the diverse styles that have emerged in the present market. Figure~\ref{fig:handicrafts} shows samples of some traditional designs. In a survey conducted by authors, 66\% of urban youth in the age group 18-25 years find these designs boring and outdated. According to the National Handicrafts Policy Report presented by All India Artisans and Craftworkers Welfare Association, artisans ``must build genuine, modern, and evolving hand crafted products that are relevant in the market" \citep{aiaca2017report}. In order to revive the economy and retain employment amongst the millions dependent on this industry, it is imperative to evolve from traditionalism and adapt to contemporary trends of the global marketplace.

In this paper, we use AI to create contemporary designs at scale for manufacturing handcrafted products. We target two ancient handicrafts of India, namely Ikat and Block Print. For Ikat, we use Conditional Adversarial Networks \citep{isola2017image} followed by Global Color Transfer to artistically color motifs. For Block Print, we use a rule-based generative approach coupled with a pruning model to create visually appealing geometric patterns. These design styles infuse a modern twist into the ancient handicrafts and improve their marketability. The main contribution of our work is providing artisans with an abundance of AI generated designs that are significantly more likeable than traditional ones. By manufacturing handicrafts featuring such designs, we hope to achieve a substantial boost of economic gains among local handicraft communities.

\section{Related Work}


Artificial Intelligence is rapidly advancing to influence multiple facets of human lives \citep{althaus2015artificial} and is also being applied in areas of social good \citep{aisocial}. For instance, significant research has been done in the fields of medicine \citep{jiang2017artificial}, agriculture \citep{dimitriadis2008applying} and disaster management \citep{tellez2009using}. There have been design interventions by NGOs and designers to revive dying Indian handicrafts \citep{harita2014design}. However, ours is the first work of applying AI to any aspect of Indian handicrafts, to the best of our knowledge.

AI techniques have been leveraged for emulating creativity \citep{boden1998creativity} and imagination \citep{mahadevan2018imagination}. Significant work has been done in the field of Generative Adversarial Networks \citep{goodfellow2014generative} and its applications for image generation \citep{elgammal2017can}, style-transfer \citep{DBLP:journals/corr/LiW16b} and image-to-image translation \citep{isola2017image}. For pattern generation, polygon splitting methods \citep{ghali2011geometry,mei2013ear} and procedural generation of tangles \citep{santoni2016gtangle,dunham2007algorithm} have been explored. There has been research on color-transfer using color-space statistics \citep{reinhard2001color,xiao2006color} and neural representations \citep{he2017neural}. Alternative approaches for color-theme-extraction using classification and clustering is mentioned in \cite{colorPalette} and \cite{kousalya2013image}. In contrast to these methods, our work generates colored motifs and patterns that are viable to be manufactured into physical products.
\vspace{0.05cm}

\section{Methodology}

In this section, we discuss the design generation approach for two handicrafts - Ikat and Block Print. 
\subsection{Ikat}

\begin{figure}[t!]
  \centering
  \includegraphics[width=0.5\textwidth]{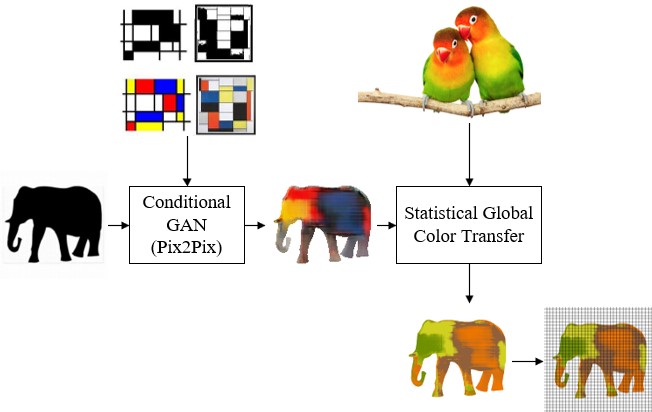}
  \caption{\small{Architectural overview of our approach to create Ikat designs}}
  \label{fig:architecture_ikat}
\end{figure}

Ikat is a dyeing technique in which the yarn is dyed prior to weaving. Popular in the state of Telangana in India, its style is in contrast to other techniques, where dyeing happens after the cloth is woven. Hence this technique creates a shading effect of different colors merging into one another. We harness this property to create designs in our approach. We use a two step process, where a black motif is first colored using a primitive color scheme, and later transformed to a color scheme based on an input inspiration. This method is described in Figure~\ref{fig:architecture_ikat}.

\subsubsection{Primitive Colorization of Motif}
\label{sec:primitive colorization}
In this step, we color a black motif using pix2pix, a Conditional Adversarial Network \citep{isola2017image}. The model is trained on a set of 1000 paintings from a famous European painter, Piet Mondrian\footnote{\url{www.wikipedia.org/wiki/Piet_Mondrian}}, and their grey-scale counterparts. The simplicity of these paintings along with the use of only primitive colors made them an ideal choice for our approach, since our model is able to learn primitive colorization of a motif from a relatively small training dataset. 


A pix2pix model uses a generator which attempts to colorize the input and a discriminator that learns to distinguish between the real paintings and the colorized images. The discriminator's output determines the loss of the generator, which the generator tries to minimize, effectively colorizing images to make them indistinguishable from real paintings. 

\subsubsection{Color Transfer from Inspiration}
\label{sec:transfer from insiration}
The primitive-colored motifs are recolored with colors of an inspiration image using a statistical approach of global color transformation. The inspiration's mean and standard deviation are imposed on a motif across the three channels of the LAB color space. Details of this transformation are out of the scope of this paper and can be read about in \citet{reinhard2001color}. Finally, the design is post-processed to a 128*128 grid that can be readily used for dyeing, as each cell is of a single color. 

Sections \ref{sec:primitive colorization} and  \ref{sec:transfer from insiration} are diagrammatically represented in Figure 2.
\subsection{Block Print}

Block Print is an ancient handicraft style practiced in the state of Rajasthan in India. In this method, patterns are carved on wooden blocks, each block is dyed with a unique color, and the blocks are then repeatedly stamped across a fabric. Due to the nature of this method, the designs have a recurring pattern and no two colors merge with each other. These characteristics are taken into consideration while creating our designs. Our approach for creating Block Print designs is a three step process of applying a rule-based approach to generate patterns, then extracting colors from an inspiration image to color them, and lastly, utilizing the pruning model to discard bad designs.

\begin{figure}[t!]
  \centering
  \includegraphics[width=0.5\textwidth]{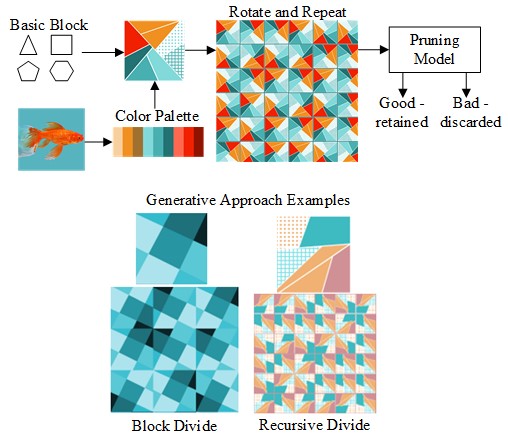}
  \caption{\small{Architectural overview of our approach to create Block Print designs}}
  \label{fig:flow_diagram_blockprint}
\end{figure}

\begin{table*}[!t]
\centering
\caption{\small Feature set used for pruning of Block Print designs}
\label{pruning-features}
\resizebox{\textwidth}{!}{%
  \begin{tabular}{p{0.1cm}p{2.8cm}p{13.3cm}}
    \toprule	
  \# & \textbf{Features} & \textbf{Description} \\
    \toprule
    1 & Area & Fraction of total area occupied by each color. \\
    2 & Darkness & The two colors occupying the most area are classified as dark or non-dark based on a threshold on Hue and Lightness properties.\\
    3 & Dullness Score & For each color, a dullness score is evaluated based on Hue, Saturation and Value properties, and then averaged out to indicate global dullness.\\
    4 & Color Harmonies & The type of Color Harmony \citep{doi:10.1002/col.10004} is based on the difference in Hue of all colors.\\
    5 & Global Contrast & For all adjacent colors, contrast is calculated using difference in their Lightness components, and then averaged out to indicate global contrast.\\
    \bottomrule
  \end{tabular}}
\end{table*}
\subsubsection{Rules for Pattern Generation}
\label{ssec:generative_approach}
Our first step is to generate several geometric patterns using the following four rules.
\vspace{-0.2cm}
\begin{enumerate}
\item Start with a geometric shape to use as the basic block such as a square, triangle or hexagon.
\vspace{-0.1cm}
\item Join any two points on distinct edges by a straight or curved line.
\vspace{-0.1cm}
\item Repeat rule 2 several times to create a block design. Based on the style in which rule 2 is repeated, we can generate different block designs. For example, we can join points that lie on block edges (Block Divide) or that recursively splits a shape in two (Recursive Divide).
\vspace{-0.1cm}
\item Rotate (optional) and repeat the block on the design board to create a unique pattern. 
\end{enumerate}
`Generative Approach Examples' in Figure~\ref{fig:flow_diagram_blockprint} shows some of the designs generated using the above rules.

\subsubsection{Color Palette Generation from Inspiration}
As a next step, designs are colored using photographs as inspirations. From all the colors of an inspiration, the ones with low prominence are removed, where prominence is determined by the color's brightness, saturation and the area occupied. The remaining set of colors is then further reduced to ten by iteratively replacing similar colors with the most prominent one, using delta-e distance \cite{sharma2005ciede2000} as a measure of color similarity. Finally, the palette is created by grouping together colors of comparable hues.

`Color Palette' in Figure~\ref{fig:flow_diagram_blockprint} is generated using this approach.

\subsubsection{Pruning Model}

To further improve the visual appeal of generated designs, we use a pruning model to discard bad designs. We create a dataset of 1100 patterns using the rules described in Section~\ref{ssec:generative_approach} and have each sample annotated by 3 judges on basis of their liking of each design. 1000 samples are used for training and 100 are reserved for testing. Using a set of rules in accordance with color theory and statistics, we obtain a feature set as indicated in Table \ref{pruning-features}. From our experiments, we observe that Gradient Boosted Machines \citep{friedman2001greedy} with Learning rate 0.3, Max Leaves as 85 and 50 Minimum Samples achieves the best performance on the test set.

Figure~\ref{fig:flow_diagram_blockprint} represents the design process for Block Print.

\section{Experimental Setup and Results}

\label{sec:issues-eval}

\subsection{Evaluation Setup}
Products manufactured with designs generated using the above approach are found to be much more visually appealing than their traditional counterparts in the present market. However, in order to get an objective analysis, we set up a system to evaluate `likeability' of the designs. In order to do this while accommodating diverse opinions, we introduce a new metric called likeability-index, where likeability-index of `x' implies that x\% of the designs are liked by at least x\% judges. Since the perception of ‘beauty’ is subjective, we got each design judged by 20 random set of annotators representing an age group of 18-25 from India.


\begin{table}[t!]
\caption{\small Likeability index for Ikat and Block Print}
\label{l-index}
\begin{subtable}
\centering
\resizebox{1\linewidth}{!}{%
  \begin{tabular}{p{7cm}p{1cm}}
    \toprule	
    \multicolumn{2}{c}{\textsc{\textbf{Ikat}}} \\
  \toprule
  Traditional designs & 28 \\
  Designs using Cycle GAN & 49 \\
  Designs using Pix2pix &  67 \\
    \bottomrule
  \end{tabular}}
\end{subtable}
\newline
\vspace*{0.1cm}
\newline
\begin{subtable}
\centering
\resizebox{1\linewidth}{!}{%
  \begin{tabular}{p{7cm}p{1cm}}
    \toprule	
    \multicolumn{2}{c}{\textsc{\textbf{Block Print}}}  \\
  \toprule
  Traditional designs & 41 \\
  Designs using generation approach & 50  \\
  Designs using generation approach + pruning & 63  \\
    \bottomrule
  \end{tabular}}
\end{subtable}
\end{table}

\subsection{Results and Analysis}

\subsubsection{Quantitative Analysis}
Our design approaches for Ikat and Block Print have a likeability index of 67 and 63 respectively. As observed in Table \ref{l-index}, on the basis of likeability-index, they significantly outperform traditional methods and baseline approaches. 

\subsubsection{Qualitative Analysis}
\begin{figure}
\centering     
\subfigure[Good]{\label{fig:good_ikat}\includegraphics[width=84mm]{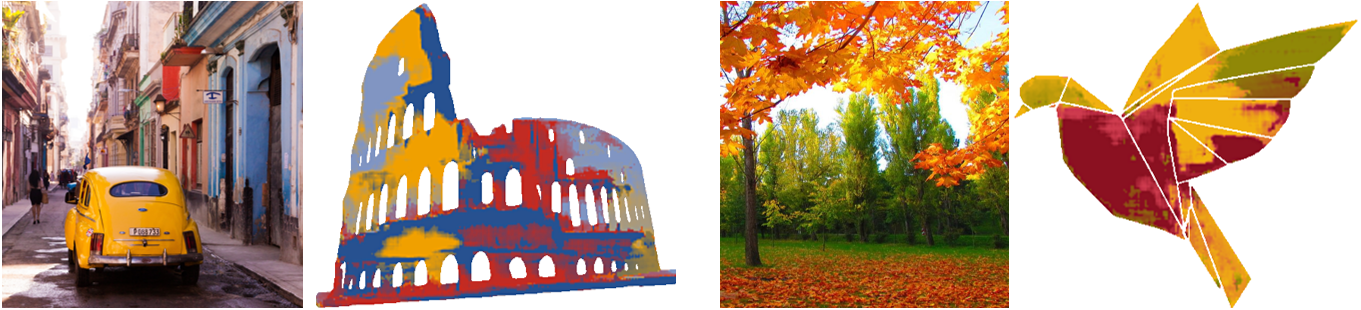}}
\subfigure[Bad]{\label{fig:bad_ikat}\includegraphics[width=32mm]{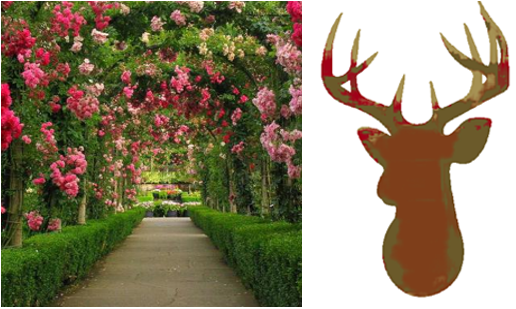}}
\caption{Designs generated for Ikat ($Inspiration \vert Motif$)}
\label{fig:examples_ikat}
\vspace{2mm}
\end{figure}

\begin{figure}
\centering     
\subfigure[Good ($Pattern \vert Inspiration$)]{\label{fig:good_blockprint}\includegraphics[width=84mm]{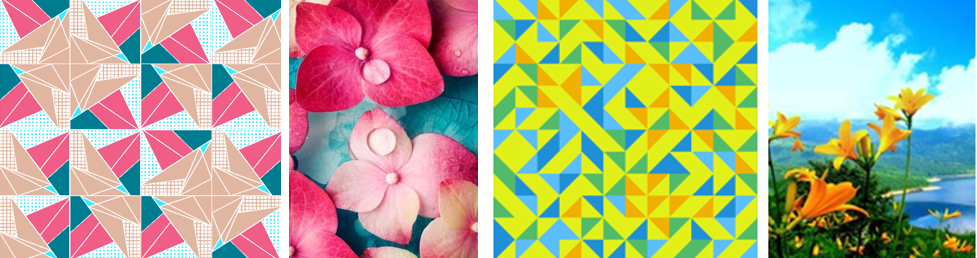}}
\subfigure[Bad (Pruned)]{\label{fig:bad_blockprint}\includegraphics[width=27mm]{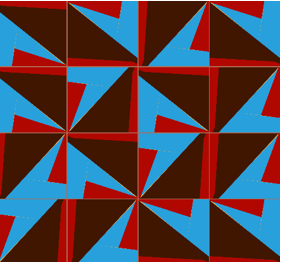}}
\hspace{2mm}
\subfigure[Bad (Not Pruned)]{\label{fig:bad_blockprint_1}\includegraphics[width=27mm]{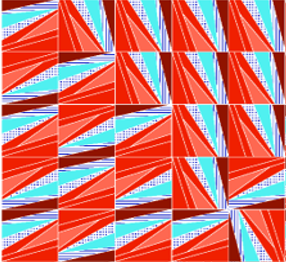}}
\caption{Designs generated for Block Print}
\label{fig:blockprint_patterns_after_pruning}
\end{figure}

Many of the designs generated for Ikat and Block Print are aesthetically appealing, as shown in Figure~\ref{fig:good_ikat} and Figure~\ref{fig:good_blockprint}. However, some of the designs are not as admirable due to factors like poor color combination or uneven color distribution, an example of which is shown in Figure~\ref{fig:bad_ikat}. For Block Print, many of the bad designs are filtered out by the pruning model. Figure~\ref{fig:bad_blockprint} shows one such example. However, it fails to remove some of the unappealing designs, as in the case of  Figure~\ref{fig:bad_blockprint_1}.
\vspace{2mm}

\section{Conclusion}

\begin{figure}
\centering     
\subfigure[Ikat]{\label{fig:ikat}\includegraphics[width=84mm]{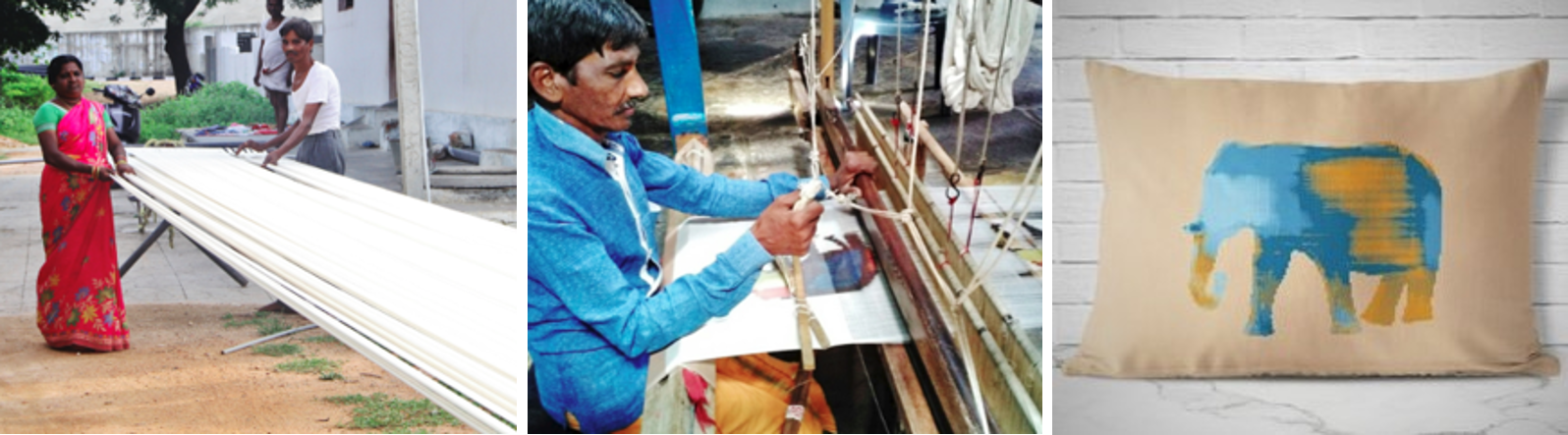}}
\subfigure[Block Print]{\label{fig:block_printing}\includegraphics[width=84mm]{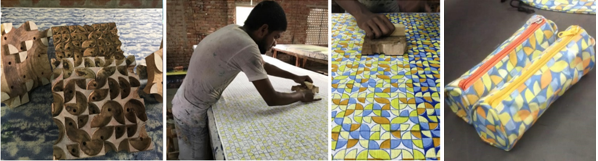}}
\caption{Artisans manufacturing handicraft products using our designs}
\label{fig:workers}
\end{figure}

In order to survive the global competition, the Indian handicrafts industry must evolve to accommodate contemporary designs at scale. We present novel techniques of creating more sought after designs for two Indian handicrafts - Ikat and Block Print. For Ikat, we use Conditional Adversarial Networks and Global Color Transfer to colorize motifs. For Block Print, we use a generative rule-based approach coupled with a pruning model to create patterns. Figure~\ref{fig:ikat} and Figure~\ref{fig:block_printing} show how weavers of Koyalagudem, Telangana and Block Print communities of Sanganer, Rajasthan used these techniques to manufacture and sell beautiful handicrafts. Designs generated by our approach are evaluated to be significantly more likeable than the conventional ones. The notable increase in marketability of the handicrafts has the potential to generate greater revenue for local artisans. However, our approach has limitations like inability of generating non-geometric designs, and at times, yielding designs with poor color distribution. We plan to address these limitations in future work.

\bibliography{references}
\bibliographystyle{icml2019}

\end{document}